\documentclass[prd,10pt,aps,twocolumn,showpacs,nofootinbib,floatfix,amssymb,amsmath,superscriptaddress]{revtex4-1}

\usepackage{slashed}
\usepackage{epsfig}

\graphicspath{ {images/} }

\newcommand{\ext}{{\mathrm{ext}}}
\newcommand{\inter}{{\mathrm{int}}}

\newcommand{\beqn}{\begin{eqnarray}}
\newcommand{\eeqn}{\end{eqnarray}}
\newcommand{\be}{\begin{equation}}
\newcommand{\ee}{\end{equation}}
\newcommand{\eq}[1]{(\ref{#1})}
\newcommand{\iI}{{({\mathrm{I}})}}
\newcommand{\iII}{{({\mathrm{II}})}}

\newcommand{\Z}{\mathbb{Z}}
\newcommand{\dg}{\slashed}
\newcommand{\onecol}[1]{\left(\begin{array}{c} #1 \end{array}\right)}
\newcommand{\twocol}[1]{\left(\begin{array}{cc} #1 \end{array}\right)}

\begin{document}

\title{Fermion zero modes in a chromomagnetic vortex lattice}

\author{M. N. Chernodub}\email{On leave from ITEP, Moscow, Russia.}
\affiliation{CNRS, Laboratoire de Math\'ematiques et Physique Th\'eorique, Universit\'e Fran\c{c}ois-Rabelais Tours,\\ F\'ed\'eration Denis Poisson, Parc de Grandmont, 37200 Tours, France}
\affiliation{Department of Physics and Astronomy, University of Gent, Krijgslaan 281, S9, B-9000 Gent, Belgium}
\author{Tigran Kalaydzhyan}
\affiliation{Department of Physics and Astronomy, Stony Brook University, Stony Brook, New York 11794-3800, USA}
\author{Jos Van Doorsselaere}
\affiliation{CNRS, Laboratoire de Math\'ematiques et Physique Th\'eorique, Universit\'e Fran\c{c}ois-Rabelais Tours,\\ F\'ed\'eration Denis Poisson, Parc de Grandmont, 37200 Tours, France}
\author{Henri Verschelde}
\affiliation{Department of Physics and Astronomy, University of Gent, Krijgslaan 281, S9, B-9000 Gent, Belgium}

\begin{abstract}
We prove the existence of zero modes of massless quarks in a background of spaghetti vacuum of chromomagnetic vortices in QCD. We find a general solution for the zero modes and show that the modes can be localized at pairs of vortices.
\end{abstract}


\date{January 17, 2013}

\maketitle

\section{Introduction}

In one of the most popular phenomenological approaches to the ground state of quantum chromodynamics, the vacuum is described as an ensemble of intertwined and entangled chromomagnetic vortices which divide the vacuum into domainlike structures~\cite{ref:NO:1978:instability,ref:NO:electricstring,ref:NO:1979:spaghetti,ref:AO:lattice,ref:AO:center}. Within each domain,
all vortices have approximately the same orientation, both in coordinate and color spaces. At larger distance scales the orientations of domains is random so that both Lorentz and color symmetries of the vacuum are restored in the infrared regime. This "spaghetti vacuum" picture makes it possible to qualitatively explain certain fundamental properties of the QCD vacuum.

For example, the confinement of quarks appears as a natural result of the randomization of the vortex domains: a single quark scatters off the vortices in each domain so that the quark's wave function acquires a random phase which varies from domain to domain. As a result of averaging over the domains, a single quark gets an infinitely large free energy which effectively forbids the existence of isolated quarks. However, a pair of a closely spaced quark and antiquark would get a correlated contribution to the phase of their common wave function, and this colorless bound state would have a finite energy in agreement with the confinement picture. As the distance between the constituents of the pair increases, the phase cancellation works less effectively so that the free energy of the system increases (linearly) with the distance. The standard interpretation of this picture is that the color charges are confined together by a linear potential. The latter is caused by a chromoelectric string spanned between the charges~\cite{ref:NO:electricstring}.

The chromomagnetic vortices are formed due to the (chromo)paramagnetic nature of the perturbative QCD vacuum which has an unstable mode towards creation of a chromomagnetic field~\cite{ref:Savvidy}. In turn, a homogeneous chromomagnetic field background itself becomes unstable --due to large chromomagnetic moment of gluon --towards squeezing of the chromomagnetic field into parallel flux tubes. The tubes are associated with chromomagnetic vortices~\cite{ref:AO:lattice}. Within each domain the vortices form a lattice structure which is similar to the Abrikosov vortex lattice in a mixed state of an ordinary type-II superconductor subjected to an ordinary external magnetic field~\cite{Abrikosov:1956sx}. The analogy between the spaghetti vacuum picture and the ordinary superconductivity may have a deeper
relation as the vacuum, within each domain, may be interpreted as a (chromo)superconducting medium from the point of the transport properties of the color charges~\cite{Chernodub:2012mu}.

Besides the qualitative explanation of color confinement, the spaghetti vacuum is tightly related to the emergence of the gluon condensate  $\langle \alpha_s G^a_{\mu\nu} G^{a,\mu\nu} \rangle$ which appears as a natural consequence of the condensation of the chromomagnetic field.

Other important features of the nonperturbative QCD vacuum, namely topological properties and chiral symmetry breaking, are captured by zero and near-zero quark modes, respectively. In order to address these properties of the QCD vacuum, a detailed study of the quark modes in the background of the gluonic spaghetti vacuum is necessary. This is the aim of the present paper.

Zero modes of fermions are known to exist in the cores of isolated vortices in a different physical context of an Abelian gauge theory~\cite{Jackiw:1981ee,Witten:1984eb,Schaposnik}. Such massless modes consist of left- and right-handed fermions which propagate back and forth along the vortex. If the fermions carry electric charge, then an applied electric field breaks the balance between oppositely moving modes thus generating  a dissipationless electric current along the vortex core. The vortex is then interpreted as a superconducting wire, hence the name "the superconducting string"~\cite{Witten:1984eb}. Thus, the presence of the fermion zero modes bound to the vortex cores is linked to emergence of  superconductivity. In our case, the fermions carry the chromoelectric charge so that the resulting superconductivity should be naturally be interpreted in a chromoelectric sense, in line with the discussions of Ref.~\cite{Chernodub:2012mu}. The dissipationless transport can be also tightly related to the phenomenon of "chiral superfluidity" in the quark-gluon plasma \cite{Kalaydzhyan:2012ut}.

In this paper we describe zero quark modes in the spaghetti picture of the QCD vacuum. Following the original formulation of Refs.~\cite{ref:NO:1979:spaghetti,ref:AO:lattice} we consider an ensemble of straight parallel chromomagnetic vortices within a single domain. A similar question was addressed in Ref.~\cite{Beri:2013fya} in a context of a $(2+1)$ dimensional gauge theory in a perturbative regime. Some of the results of Ref.~\cite{Beri:2013fya} --for example, the density of the zero mode in transverse plane --turn out to be qualitatively similar to the properties of the exact $3+1$ dimensional zero modes discussed in this paper.

It is worth mentioning that the interest in the fermionic zero modes in a background of (non-Abelian) Abrikosov-like vortex configurations is not exclusively limited to the spaghetti picture of QCD vacuum. The non-Abelian vortex structures coupled to light fermionic fields appear in different physical contexts. For example, the electroweak vacuum in a background of sufficiently strong magnetic field may develop a vortex lattice ground state immersed in the $W$-boson condensate~\cite{ref:AO:EW}. Recently it was found that this ground state is characterized by anisotropic superconductivity along the vortices~\cite{ref:EW:superconductivity}. A similar phenomenon may occur also in QCD caused by a suggested $\rho$-meson condensation~\cite{ref:QCD:superconductivity} for much lower and possibly experimentally reachable magnetic fields~\cite{Skokov:2009qp}. In the condensed matter context, a strong magnetic field was suggested to induce
a "reentrant" superconductivity via a formation of
an inhomogeneous vortex ground state in strong type-II superconductors~\cite{ref:Tesanovic} and as well as
in ferromagnetic superconductors~\cite{ref:Olesen:recent}.

The structure of the paper is as follows. In Sec.~\ref{sec:vortex:lattice} we describe the spaghetti vacuum and discuss the coupling of the chromoelectric vortices to massless fermions. In Sec.~\ref{eq:sec:exact} we construct a system of zero-mode equations, and find a general solution for normalizable and gauge-covariant zero modes. In Sect.~\ref{eq:sec:exact:boundary} we discuss a relation between the localization properties of the zero modes and boundary conditions. Our last section is devoted to conclusions.

\section{Vortex lattice and fermions}
\label{sec:vortex:lattice}

\subsection{Vortex lattice solution}

First, we briefly outline the structure of the gluon field within each domain of spaghetti vacuum~\cite{ref:AO:lattice}. The full solution for the gauge field is obtained by a proper embedding of a $SU(2)$ solution in a subgroup of the full $SU(3)$ gauge group. Without loss of generality, we choose the chromomagnetic field $B > 0$ to lie along the third $x^3$ direction, so that all nonzero components of the gluon field correspond to the transversal $(x^1,x^2)$ plane. The chromomagnetic field $B$ is encoded into the diagonal gluon components of the $SU(2)$ solution,
\beqn
	 A^3_1&= -\frac{B}{2}x^2,\quad A^3_2=\frac{B}{2}x^1,\label{def:gauge}\\
	 A^1_\mu&\mp iA^2_\mu= \sqrt{2}W^\pm_\mu,
\eeqn
while the off diagonal gluon field,
	\be
	W^-_1=iW^-_2=(W^+_1)^*=-i(W^+_2)^*=W/2,
	\ee
should satisfy the following classical equation of motion:
	\be
	\left[\left(\partial_1 - i \partial_2 \right) + \frac{gB}{2} \left(x^1-ix^2\right)\right]W = 0.
	\label{abr}
	\ee

From now on it is convenient to use complex dimensionless variables for the coordinates in the transverse plane,
\be
z=\frac{x^1+ix^2}{L_B},\quad \bar z=\frac{x^1-ix^2}{L_B},
\label{def:dimless}
\ee
and write all dimensional quantities in terms of the magnetic length
\beqn
L_B = \sqrt{\frac{2 \pi}{g B}}\,.
\label{eq:LB}
\eeqn

Then the solution of Eq.~\eq{abr} for the Abrikosov lattice of the chromomagnetic vortices can be expressed via a holomorphic function $\omega$:
\beqn
W(z,\bar z)&=
\omega(\bar z)e^{-\frac{\pi}{2}z\bar z}.
\label{Wh}
\eeqn
The choice of the exponent of solution~\eq{Wh} ensures
that we have one vortex per unit of transverse surface.

        \begin{figure}[!htb]
		\begin{center}
	 \includegraphics[scale=0.35]{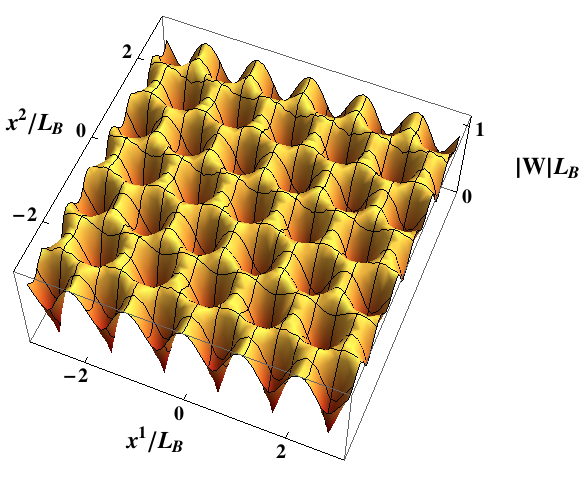}
	 \caption{The condensate $W$ -- given by Eqs.~\eq{Wh}, \eq{def:omega} and \eq{eq:nu} -- is a periodic lattice of single winding (anti-)vortices as a function of normalized coordinates in the transversal plane.}\label{figW}
	 \end{center}
	 \end{figure}
	
We will assign to $\omega$
a well-known expression for a vortex lattice~\cite{ref:footnote1}:
	 	\be
	 	\omega(\bar z)=C_0\theta_1\left(\nu {\bar z},q\right)e^{\frac{\pi}{2}\bar z^2},
		\label{def:omega}
	 	\ee
where $\theta_1$ is the Jacobi $\theta$ function, and $C_0 = \sqrt{2 \pi} c_0/g$ with $c_0 \approx 1$ \cite{ref:AO:lattice}.
The number $q$ is defined by
	 	\be
	 	q=e^{i\pi\tau}\,,
		\ee
and for a hexagonal (equilateral triangular) or square lattice we have the following choices, respectively:
		\be
		\tau=e^{i\frac{\pi}{3}},\qquad\tau=i.
	 	\ee
The hexagonal lattice corresponds to the minimum of the free energy of the system~\cite{ref:AO:lattice}, and
the parameter $\nu$ takes the following value:
\beqn
\nu=\sqrt{{\rm {Im}}( \tau)}\,.
\label{eq:nu}
\eeqn

	  	The important point in this choice is that now the zeroes of $\omega$ are now given by
	 	\be
	 	X_{m,n}=\frac{2}{\nu}(m+n\tau),\quad \bar X_{m,n}=\frac{2}{\nu}(m+n\tau^*),
		\label{defX}
	 	\ee
	 	for all integer and half-integer values of $m$ and $n$. This is clearly visible in the plot of the condensate $W$, Fig.~\ref{figW}.

\subsection{Coupling to fermions}

The gluon field couples to the fermions through a massless Dirac operator $\dg{D}$. It is convenient to split the chirality matrix $\gamma^5$ into {\it internal} chirality $\gamma^\inter$ and {\it external} chirality $\gamma^\ext$ matrices (the notations are borrowed from Ref.~\cite{ref:Callan:Harvey}):
	\be
	\gamma^\ext=i\gamma^1\gamma^2\,,\quad \gamma^\inter=\gamma^0\gamma^3\,,\quad\gamma^5=\gamma^\ext\gamma^\inter\,,
	\ee
and introduce the corresponding projectors:
	\be
	\Gamma^\pm_\ext = \frac{1\pm \gamma^\ext}{2}\,,
	\qquad
	\Gamma^\pm_\inter = \frac{1\pm \gamma^\inter}{2}\,.
	\label{eq:Gamma:pm}
	\ee
The external projectors $\Gamma^\pm_\ext$ single out the spin-up and spin-down fermion states (as characterized with respect to the direction on the magnetic field $B$) while the internal projectors $\Gamma^\pm_\inter$ select the left- and right-handed fermions (so that the left- and right-handed fermions with the same spin state move in opposite directions along the vortex string). All matrices
$\Gamma^\pm_\ext$, $\Gamma^\pm_\inter$ and $\gamma^5$ commute with each other so that they have a common eigenbasis and the same energy spectrum.

In order to find the zero modes, we need to solve the Dirac equation,
    \be
    \dg{D} \Psi = 0.
    \label{eq:Dirac}
    \ee
In the light-cone coordinates $x_\pm = (x^0\pm x^3)/2$ the Dirac operator in Eq.~\eq{eq:Dirac} can be rewritten as follows,
    \be
	\dg{D}(x_\pm,z,{\bar z})=\gamma^0\Gamma^+_\inter \partial_{x_+} + \gamma^0 \Gamma^-_\inter \partial_{x_-} + \dg{D}_\perp(z,{\bar z})\,,
\label{eq:dg:D}
	\ee
where $\dg{D}_\perp$ is an operator proportional to the transverse $\gamma^1$ and $\gamma^2$ matrices which depends on transverse coordinates $x^1$ and $x^2$ only.

The spaghetti background~\eq{Wh} is independent on the longitudinal ($x_3$) and time ($t$) coordinates, so that any zero-mode 
solution of the Dirac equation~\eq{eq:dirac} can be written as follows:
\beqn
\Psi({\boldsymbol x}, x^0) = e^{- i \varepsilon x^0 + k_3 x^3} \psi(x_1,x_2)\,,
\label{eq:Psi}
\eeqn
where $\psi(x_1,x_2)$ is a solution of the transverse Dirac equation corresponding to a zero mode in the transverse plane,
    \be
     \dg{D}_\perp(z,{\bar z})\psi(z,{\bar z}) \equiv 0.
    \label{eq:dirac}
    \ee
The spectrum of the full zero-mode solution~\eq{eq:Psi} corresponds to the massless modes,
\beqn
\varepsilon = \pm |k_3|\,,
\label{eq:varepsilon}
\eeqn
propagating along the core of the vortex with the speed of light. The modes are double degenerate both in terms of the spin index (spin-up and spin-down modes) and in terms of the direction of motion (up-moving and down-moving modes).

Below we will discuss only the solutions to the two dimensional Dirac equation in the transverse plane~\eq{eq:dirac}.

We can rewrite the components of the Dirac operator,
\beqn
	\dg{W}^- & = & W\gamma_1\Gamma^-_\ext,\qquad \dg{W}^+=W^*\gamma_1\Gamma^+_\ext,\\
	\dg{A} & = & i\frac{B}{2}\gamma^1\left(z\Gamma^-_\ext - \bar z\Gamma^+_\ext\right),\\
	 \dg{\partial} & = & \gamma^1\left(\Gamma^-_\ext \partial+\Gamma^+_\ext \bar\partial\right),
\eeqn
so that the transverse Dirac operator takes the following form:
	\be
	\dg{D}_\perp=\dg{\partial}-ig\dg{A}=D_- \gamma_1\Gamma^-_\ext + D_+\gamma_1\Gamma^+_\ext\,,
	\label{eq:Dirac1}
	\ee
and the operators
\beqn
D_-&=\twocol{2\partial_{\bar z}+\frac{\pi}{2} z&0\\-\frac{i g}{\sqrt{2}}W(z, {\bar z}) & \quad 2\partial_{\bar z}-\frac{\pi}{2}z},\\
D_+&=\twocol{2\partial_z-\frac{\pi}{2} z& \quad - \frac{i g}{\sqrt{2}}W^*(z, {\bar z})\\0&2\partial_z+\frac{\pi}{2}\bar z},
\eeqn
correspond to the spin-down and spin-up projections, respectively.
	
Clearly, the orthogonal eigenspaces of the projectors $\Gamma^\pm_\ext$ coincide with the eigenspaces of the operator $\gamma_1\dg{D}_\perp$ so that we can treat both of them independently. The spin-up and spin-down solutions, respectively,
	\be
	\Gamma^-_\ext \psi=\psi^- \equiv \onecol{\psi^-_1\\\psi^-_2},
	\qquad
	\Gamma^+_\ext \psi=\psi^+ \equiv  \onecol{\psi^+_1\\\psi^+_2},
	\label{def:xi:zeta}
	\ee
are essentially equivalent to each other, so we restrict our attention to the solution $\psi^-$ only.

In the $\Gamma^-$ subspace this equation reduces to
	\be
\twocol{2\partial_{\bar z}+\frac{\pi}{2} z&0\\-i\frac{g}{\sqrt{2}}W(z, {\bar z})&\quad 2\partial_{\bar z}-\frac{\pi}{2}z}\onecol{\psi^-_1 \\ \psi^-_2} =0\,.
\label{eq:psiminus}
	\ee

    Note that an up- or down-moving zero mode does have nonzero energy, but its contribution to the energy stems solely from the longitudinal kinetics. The only true zero energy configurations are static in the $x^3$ direction, with $k_3 = 0$ in Eq.~\eq{eq:varepsilon}. It is exactly the purely transversal configuration we want to find in the next section by solving Eq.~\eq{eq:dirac}.

The problem of solving the Dirac equation (\ref{eq:dirac}) in the transversal plane was also considered in Ref.~\cite{Beri:2013fya} in a slightly different setting. The authors found a zero mode in $2+1$ dimensions in a limit of a weak gluon field,
\be
| W(x,y) | L_B\ll 1,
\label{eq:weak:field}
\ee
where $L_B$ is the magnetic length~\eq{eq:LB}. By treating the gluon field $W$ as a perturbation, the Hamiltonian was diagonalized in terms of the unperturbed eigenstates (\ref{eq:psiminus}) and their conjugates, which are bounded only in their upper and, respectively, lower components. The Brillouin zone of these perturbative solutions was revealed to hold two Dirac points which correspond to zero-mode solutions. Reference~\cite{Beri:2013fya} gives a very interesting treatment of the topology of gapped solutions in two spatial dimensions in a system of two fermion flavors. Below we describe the exact nonperturbative solutions for zero modes beyond the weak-field approximation~\eq{eq:weak:field}.

\section{Exact zero modes}
\label{eq:sec:exact}

In this section we find a most general solution to Eq.~(\ref{eq:psiminus}) without imposing boundary conditions which are closely related to the topological properties of the zero modes. The boundary conditions --which parameterize both independent solutions of Eq.~(\ref{eq:psiminus}) --will be discussed in the next section.

\subsection{First solution}

To find a first solution, we will construct a zero-mode solution for a reduced version of the system (\ref{eq:psiminus}) by setting one of the components, either $\psi_1^-$ or $\psi_2^-$, to zero.

We conclude quickly that $\psi_2^-= 0$ immediately implies the triviality of the full solution.
Indeed, the $W$ condensate~\eq{Wh} is a nonvanishing function almost everywhere so that Eq.~(\ref{eq:psiminus}) implies $ \psi_1^-= 0$.

One can alternatively set the upper component to zero, $\psi_1^{\iI,-} = 0$. Then the lower component should satisfy the following equation:
	\be
	 \left(\partial_{\bar z}-\frac{\pi z}{4}\right)\psi_2^{\iI,-}=0\,,
	\ee
implying that the first set of solutions of Eq.~(\ref{eq:psiminus}) has the following form:
	\be
	\psi_1^{\iI,-}=0\,,\qquad
	\psi_2^{\iI,-} (z,\bar z)=W(z,\bar z)^{-\frac{1}{2}}
	\,,
	\label{eq:sol1}
	\ee
where the function $W(z,\bar z)$
is given in Eqs.~\eq{Wh} and \eq{def:omega}.

This solution has a large degeneracy as it can be multiplied by an arbitrary holomorphic function $h_1(z)$ and it still will be a zero-mode solution.
It is important to notice that the holomorphic nature of the function $h_1$ is a severe constraint. Rather than being a free function in each point of the transversal plane, it is determined by its behavior on a finite line, say at the boundary of the condensate and by its behavior around the poles.

 \begin{figure*}[thb]
	 \begin{center}
	 \includegraphics[scale=0.27]{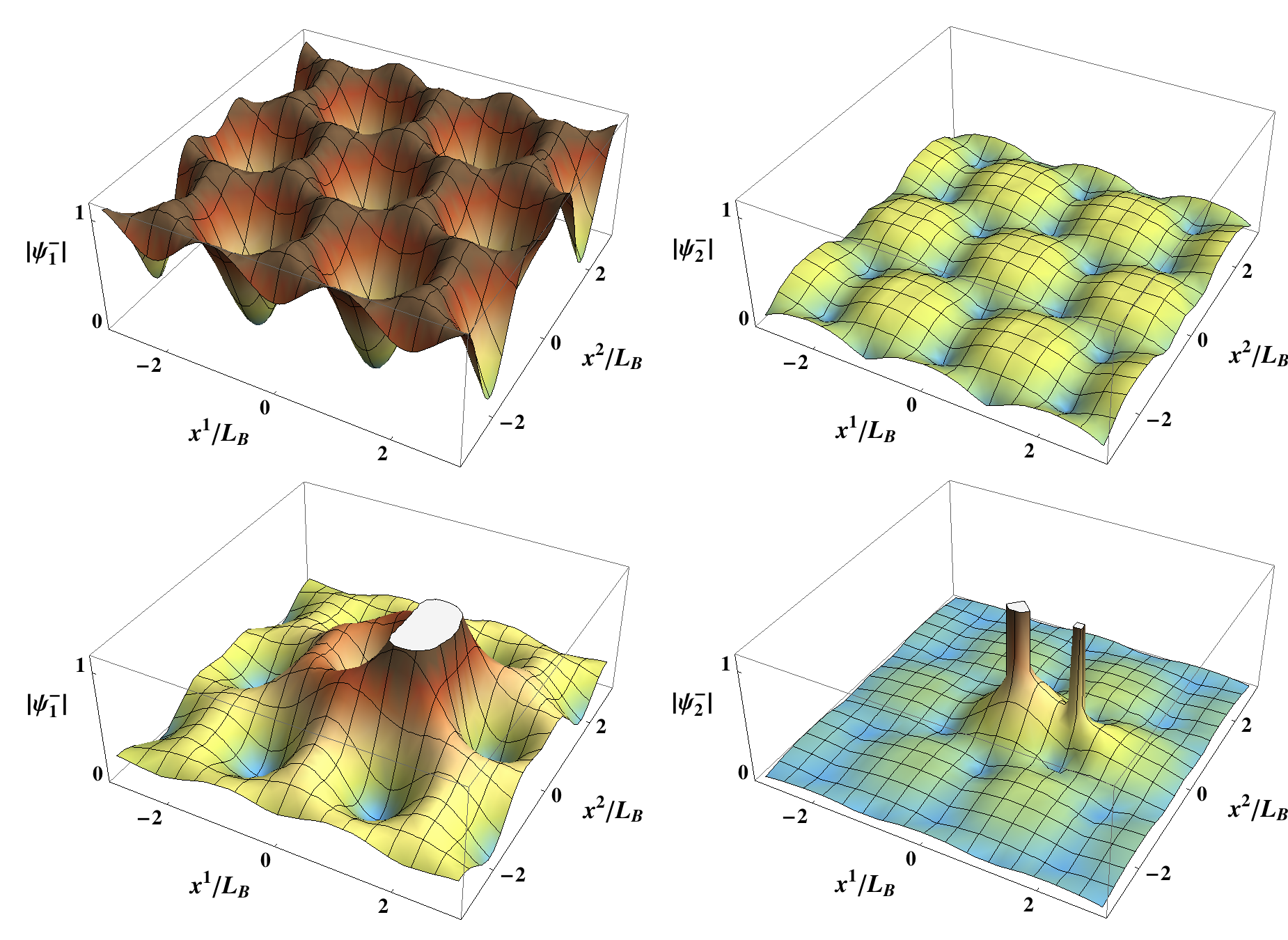}
\caption{Absolute values of the zero-mode components
$\psi^-_1$ (left) and $\psi^-_2$ (right) of the general solution $\psi^-$, Eq.~\eq{eq:genexp}
with $\mu(z)$ given in Eq.~\eq{eq:mu}.
The upper
panel
corresponds to a periodic solution with $h_1=0$ and $h_2=1$. The periodic lattice has only half the periodicity of the condensate shown in Fig.~\ref{figW}. The lower
panel visualizes the localized solutions with
$h_1=0$ and with the holomorphic function $h_2(z)$ given in Eq.~(\ref{twopoles}). }
     \label{fig:xi:zeta}
	 \end{center}
	 \end{figure*}

\subsection{Second solution}

We will now look for a second solution to Eq.~(\ref{eq:psiminus}) considering a nonzero function $\psi_1^-(z,\bar z)$ from the very beginning. From Eqs.~(\ref{abr}) and (\ref{eq:psiminus}) we see that the functions $W$ and $\psi_1^-$ should obey a similar Abrikosov equation. We define
	\be
	\psi_1^{\iII,-}(z,\bar z) = \chi \mu^2(z)e^{-\frac{\pi}{4}z\bar z},
	\label{eq:psi:1}
	\ee
where $\mu(z)$ a scalar
holomorphic
function which is vanishing at even positions of the vortex cores~\eq{defX}:
\beqn
\mu(X_{m,n}) = 0\,,\qquad m,n \in \Z\,.
\label{eq:mu:0}
\eeqn
In Eq.~\eq{eq:psi:1} $\chi$ is a constant polarization
two-component spinor.
The exact form of the polarization spinor $\chi$ is irrelevant for our discussion of this transversal
two
dimensional problem, and therefore we omit it below. This spinor can be expanded in the basis of two eigenfunctions of the internal chirality operator $\Gamma^\pm_\inter$ given in Eq.~\eq{eq:Gamma:pm}, corresponding to the states moving up and down along the longitudinal direction.

To write the condensate in a convenient way, we use an antiholomorphic function $\omega(\bar z)$ defined in Eq.~\eq{def:omega} so that the condensate $W(z, {\bar z})$, Eq.~\eq{Wh}, satisfies the equation of motion~(\ref{abr}). One way to obtain similar periodic behavior for $\mu(z)$ while keeping $\psi_1^-$ finite, is to put
	\be
	\mu(z)=\omega\left(\frac{z}{2}\right),\label{eq:mu}
	\ee
at the expense of reducing the periodicity and quadrupling the surface of a unit cell. Equation~\eq{eq:mu} is a minimal single-valued choice for the function $\mu$ consistent with condition~\eq{eq:mu:0}.
This choice implies that $\psi_1^-$ has one double zero per four unit cells, or equivalently, has a branch point behaving like $z^{\frac{1}{2}}$ in each
 cell. This is, of course, in direct correspondence to the $\mathbb{Z}_2$ behavior of each vortex. To keep the analytical expression manageable, however, it is useful to group four cells and avoid fractional branch cuts
inside the cells.
		
We now are left with the calculation of the lower component of the second zero-mode solution $\psi_2^{\iII,-}$. We rewrite it using the (anti-)holomorphic functions $\omega(\bar z)$ and $\mu(z)$, and a new function $f(z,\bar z)$:
	\be
	\psi_2^{\iII,-}(z,\bar z)=f(z,\bar z)e^{\frac{\pi}{4}z\bar z}\mu^2(z) .
	\label{zeta}
	\ee
	The remaining equation in (\ref{eq:psiminus}) becomes then
	\be
	\partial_{\bar z}f(z,\bar z) = - \frac{i g}{2\sqrt{2}}\omega(\bar z)e^{-\pi z\bar z}\,.
	\label{zetaeq}
	\ee

In the Appendix we derive a convenient expansion~(\ref{w2}) of the Jacobi $\theta_1$ function which enters the function $\omega$ in Eq.~\eq{def:omega}. Using this expansion we set
 	\be
	f(z,\bar z)=\sum_{m,n \in \Z} e^{-\frac{\pi}{2}\vert X_{m,n}\vert^2}f_{m,n}(z,\bar z)\label{fexp},
    \ee
so that the equation (\ref{zetaeq}) becomes, term by term, as follows:
    \be
	\partial_{\bar z}f_{m,n}= -i\frac{gN C_0}{2\sqrt{2}}e^{\frac{i\pi}{8}}(\bar z-\bar X_{m,n})e^{-\pi\bar z(z-X_{m,n})}.	
	\ee
   We find now for all integer $n$ and $m$:
	\beqn
	f_{m,n}(z,\bar z)& =& \frac{igNC_0}{2\sqrt{2}\pi}e^{\frac{i\pi}{8}}\left(\frac{1}{\pi}+\vert z-X_{m,n}\vert^2\right)\nonumber \\
	& & \times \frac{e^{-\pi\bar z(z-X_{m,n})}}{(z-X_{m,n})^2}
\,,
\label{eq:f:mn}
	\eeqn
	and to avoid divergences we need to set in Eq.~\eq{eq:f:mn} the integration constant, which is an arbitrary holomorphic function, to zero.
 From Eqs.~(\ref{fexp}) and (\ref{zeta}) we then get both components of the second solution:
	\beqn
	\psi_1^{\iII,-}(z,\bar z) & = & \mu^2(z)e^{-\frac{\pi}{4}z\bar z}\,,
	\label{eq:sol2:1}\\
	\psi_2^{\iII,-}(z,\bar z) & = & \frac{igNC_0}{2\sqrt{2}\pi}e^{\frac{i\pi}{8}}\mu^2(z) e^{-\frac{\pi}{4}z\bar z}
	\label{eq:sol2}\\
	& & \hskip -19mm \times \sum_{m,n \in \Z} \left(\frac{1}{\pi}+\vert z-X_{m,n}\vert^2\right)
	\frac{e^{-\frac{\pi}{2}\vert X_{m,n}\vert^2+\pi {\bar z}X_{m,n}}}{(z-X_{m,n})^2}\,.
	\nonumber
	\eeqn
For a finite zero-mode solution the holomorphic function  $\mu(z)$ should satisfy Eq.~\eq{eq:mu:0}. A minimal choice for the function $\mu$ is given in~\eq{eq:mu}. Similarly to the first solution~(\ref{eq:sol1}), we can multiply both $\psi_1^{\iII,-}$ and $\psi_2^{\iII,-}$ by an arbitrary holomorphic factor $h_2(z)$ and we will still have a solution.
	
It is interesting to note that the periodicity of the zero-mode solution with the choice~(\ref{eq:mu}) is doubly reduced compared to the periodicity of the original $W$ condensate~\eq{Wh}. This property is an inevitable consequence of the fact that no fractional branch cuts are introduced in our construction and thus the $\mathbb{Z}_2$ behavior of the solution is absent, while the fundamental $SU(2)$ fermions on a periodic one-flux-quantum lattice would require such behavior.

\subsection{General solution}
	
A general zero-mode solution $\psi^-$ is a sum of the first solution~\eq{eq:sol1} and the second solution, Eqs.~\eq{eq:sol2:1} and \eq{eq:sol2}, multiplied by arbitrary holomorphic functions $h_1(z)$ and $h_2(z)$, respectively:
	\beqn
	\psi^- & (z,{\bar z}) = & h_1(z) \psi^{\iI,-}(z,{\bar z}) + h_2(z) \psi^{\iII,-}(z,{\bar z}) \nonumber \\
	& \equiv & {h_2(z)\psi^{\iII,-}_1 \choose h_1(z) \psi^{\iI,-}_2+h_2(z)\psi_2^{\iII,-} }\,.
	\label{eq:genexp}
	\eeqn
Certain interesting examples of upper and lower components of the zero-mode solution $\psi^-$ are visualized in Fig.~\ref{fig:xi:zeta}.

\section{General solution and boundary conditions}
\label{eq:sec:exact:boundary}	

A fixing of a concrete form of the holomorphic functions $h_{1,2}(z)$ in the most general zero-mode solution~\eq{eq:genexp} corresponds to choosing specific boundary conditions for our solution, and vice versa. The boundary conditions can be divided roughly into two classes
based on their behavior towards the boundary of the system: the ones which corresponds to a periodic lattice of zero modes (a relevant example is given in the upper panel of Fig.~\ref{fig:xi:zeta}) and the localized solutions which give zero modes with finite overall
fermion number
(e.g., the one which is is given in the lower panel of Fig.~\ref{fig:xi:zeta}), and therefore converging sufficiently fast to zero at spatial infinity.
	
In addition to the boundary conditions at spatial infinity, one may also impose the conditions at the vortex cores where the gluon condensate is vanishing.
Notice that the $\mathbb{Z}_2$ part of the gauge group is not fixed by the gluon condensate because the gluons are transforming in adjoint representation of the gauge group. However, the fermion degrees of freedom are sensitive to the $\mathbb{Z}_2$ subgroup, so that each vortex core may correspond -- depending on our choice of the boundary conditions -- to a square root branch point exhibiting a multivalued behavior. We will see that this property is essential for finding all possible zero-mode solutions.
	
	At any vortex core (located, for example, at $z = 0$), a zero-mode solution has the following behavior:
	\be
	\vert\psi^-\vert^2\sim (\ldots)\frac{\vert h_1(z)\vert^2}{\vert z\vert}+(\ldots)\vert h_2(z)\vert^2\,
	\ee
where the ellipsis correspond to regular functions at $z = 0$. In order for the mode to be integrable in the transversal plane, a singularity of the function $h_1(z)$ at $z = 0$ should be less severe than $z^{-\frac{1}{2}}$. Moreover, the function $h_2(z)$ must converge faster than $z^{-1}$ as $z \to 0$. An additional constraint for the zero mode is coming from the fact that the nontrivial winding (monodromy) around the vortex cores should belong to the (unbroken) center $\mathbb{Z}_2$ of the gauge group. Therefore, half-integral powers of $z$ are allowed, and by consequence the function $h_1(z)$ is, in fact, analytical at $z = 0$, contrary to the function $h_2(z)$, for which the analyticity at $z = 0$ is not required, in general.
	
	Another important consequence
of the integrability of the zero-mode solution
is that all singularities of the zero mode should correspond to nontrivial winding only which, in turn, can only occur in the vortex cores. Therefore, {\it all singularities in a general zero-vortex solution are pinned to the background lattice of the vector boson condensate.} For a fixed background, this requirement severely restricts possible zero-mode solutions. In particular, it prohibits a continuous transition of
zero modes localized at different vortices. In a vortex liquid of real dynamical QCD -- where the background lattice is by no means constrained to the vortex lattice -- the mentioned restriction is a much less severe constraint.
	
Let us now consider periodic solutions. Because both independent zero-mode solutions, $\psi^{\iI,-}$ in Eq.~(\ref{eq:sol1}) and $\psi^{\iII,-}$ in Eq.~(\ref{eq:sol2}), are already periodic, any periodic set of the coefficient functions $(h_1,h_2)$ should give us a proper periodic solution. Since we already know that the function $h_1$ should be analytical, then the requirement of periodicity singles out a constant solution for this function. On the other hand, the function $h_2$ is not restricted to be a constant, and in this case it must have square root singularities according to our previous discussion. The only ambiguity allowed by the residual $\mathbb{Z}_2$ degeneracy appears in the sign of the fermion wave function, so that the function $h_2(z)$ must be the square root of a periodic single-valued function with simple poles.

Thus, in order to describe a periodic zero-mode solution, the coefficient functions $h_1$ and $h_2$ should satisfy the following constraints:
	\be
	h_1(z)={\mathrm{const}}.\,,\qquad h_2(z)=\sqrt{s(z)}\,,
	\ee
where the poles and simple zeroes of the function $s(z)$ must coincide with vortex cores of the background lattice. The function $s(z)$ must therefore be an elliptic function on a sublattice of the original vortex, and it must have simple poles only.
A solution with $h_2(z) \equiv {\mathrm{const}}$ corresponds to a periodic solution found in \cite{Beri:2013fya}.
An illustration of the periodic solution is given in the upper panel of Fig.~\ref{fig:xi:zeta}.

There are also zero-mode solutions which are localized around a single vortex or a compact group of vortices.
The density of the localized zero modes tends to zero at spatial infinity of the transverse plane, so that $h_1 \equiv 0$ for the localized modes.
In order for the zero mode to be normalizable in the vicinity of the vortex cores,
the function $h_2$ still needs to possess at least one singular square root cut at a vortex position.
However, in this simplest case the solution turns out to non-normalizable in the infrared sense. Thus, a general localized solution is given by Eq.~\eq{eq:genexp} with the following set of the coefficient functions:
\beqn
	h_1(z)\equiv 0\,,\qquad h_2(z)=\prod_{i = 1}^N\sqrt{\frac{1}{z-z_i}}\,,\quad N\geq 2\,,
\eeqn
where the $z_i$ correspond to a certain arbitrary set of vortex positions in the transverse plane.
The vortex positions coincide with the zeros of the gluon condensate~\eq{defX}.

The
 simplest example of the localized mode is given by the following expression:
\beqn
	\psi^- (z, {\bar z})
= \frac{C_-}{\sqrt{z\left(z-X_{\frac{1}{2},0}\right)}}{\psi_1^{\iII,-} (z, {\bar z}) \choose \psi_2^{\iII,-} (z, {\bar z}) }\,,
	\label{twopoles}
\eeqn
where $C_-$ is a normalization constant. According to Eq.~\eq{defX}, the mode is localized at the vortices located at points $z_1 = X_{0,0} \equiv 0$ and $z_2 = X_{\frac{1}{2},0} \equiv 1/\nu$ of the transverse plane.
An illustration of the localized zero-mode solution~\eq{twopoles} is given in the lower panel of Fig.~\ref{fig:xi:zeta}.

\section{Conclusions}

In this paper we have discussed properties of massless quarks in QCD in the framework of the spaghetti vacuum of chromomagnetic vortices. We have shown that the chromomagnetic vortices support zero modes of fermions propagating along the vortices with the speed of light. The modes are double degenerate both in terms of their spin polarizations and directions of motion.

Following the original works on the spaghetti picture~\cite{ref:AO:lattice,ref:AO:center}, we have concentrated on the properties of the quark modes in a regular background of a chromomagnetic vortex lattice. We have given two independent zero-modes solutions, in Eq.~\eq{eq:sol1} and in Eqs.~\eq{eq:sol2:1} and \eq{eq:sol2}. A general expression for a zero mode is provided in Eq.~\eq{eq:genexp}. We have explicitly discussed two limiting cases of these zero modes: (i) a zero-mode solution which is periodic in the transverse plane and (ii) a nonperiodic zero mode localized at a pair of chromomagnetic vortices. The periodic mode has a half reduced periodicity of the underlying vortex lattice. Both periodic and localized zero modes are visualized in Fig.~\ref{fig:xi:zeta}.

The real quantum spaghetti vacuum is expected to be described by a liquid of chromomagnetic vortices which lacks the spatial periodicity at large distances~\cite{ref:NO:1979:spaghetti,ref:AO:lattice,ref:AO:center}. Although we have discussed the fermionic solutions in the background of the regular periodic vortex structures, we expect that the zero modes should exist beyond the vortex lattice picture. Indeed, a general class of our zero-mode solutions does not rely on the periodicity of the underlying vortex vacuum. The only physical constraint coming from our analysis is a requirement for a  fermion zero mode to be shared by at least two vortices. This constraint comes as a natural consequence of the presence of the $\Z_2$ center symmetry of the gluonic vortex solutions embedded into the larger $SU(3)$ color group.

Zero fermion modes localized at vortex solutions may support dissipationless transport of charge associated with these fermions. The celebrated example of this phenomenon is given by the superconducting string in an Abelian gauge theory~\cite{Witten:1984eb}. The presence of the discussed zero-mode solutions in the spaghetti vortex background may imply the existence of superconducting chromoelectric currents propagating along the chromomagnetic vortex cores, in agreement with the recent suggestion of Ref.~\cite{Chernodub:2012mu}.

\acknowledgments
	The work of M.N.C. was partially supported by Grant No. ANR-10-JCJC-0408 HYPERMAG of Agence Nationale de la Recherche (France) and by the Chinese-French Scientific Exchange program Cai Yuanpei. The work of J.V.D. was supported by a grant from La Region Centre (France). The work of T.K. was supported by the U.S. Department of Energy under Contract No. DE-FG-88ER40388. The authors are grateful to David Tong for useful discussions.

\appendix

\section{\\[2mm] A convenient expansion of the Jacobi $\theta$ function}

To calculate the lower component of the second solution $\psi_2^{\iII,-}$ we should integrate Eq.~(\ref{zetaeq}) which includes, via Eq.~\eq{def:omega}, a Jacobi $\theta$ function. A naive way to make this final integration would be to replace the $\theta_1$ in $\omega$ by its standard series expansion and then integrate it on a term-by-term basis. However, this way of integration proves to be impossible due to noninterchangeability of integration and summation in this particular case. In order to circumvent this problem we use an alternative expansion of the function $\theta_1$.
		
In this appendix we show that a holomorphic function
\be
\tilde\omega(\bar z)=N\,C_0\, e^{\frac{i\pi}{8}}\sum_{m,n} (\bar z-\bar X_{m,n})e^{-\frac{\pi}{2}\vert X_{m,n}\vert^2+\pi\bar z X_{m,n}},
\label{w2}
\ee
is exactly equal to the function $\omega(\bar z)$ in  (\ref{def:omega}) for an appropriate choice of the normalization constant $N$.
	
We would first like to prove that the zeroes of $\omega(\bar z)$ are also zeroes of  $\tilde \omega(\bar z)$.
Put $\bar z_0=\frac{1}{2}X_{k,l}$ with $k,l\in\mathbb{Z}$
\begin{multline}
\sum_{m',n'} (\bar X_{m',n'}-\bar z_0)e^{-\frac{\pi}{2}\vert X_{m',n'}\vert^2+\pi\bar z_0 X_{m',n'}}e^{i\pi Im(\bar X_{kl}X_{m'n'})}\\
=\sum_{m,n} (\bar z_0-\bar X_{m,n})e^{-\frac{\pi}{2}\vert X_{m,n}\vert^2+\pi\bar z_0 X_{m,n}}\,.
\end{multline}
Here we have put $m'=k-m$
and
$n'=l-n$
such that ${\rm{Im}}(\bar X_{kl}X_{m'n'})$ is always an even integer. Then, we find:
\be
\tilde \omega(\bar z_0)=0\,.
\ee

The above property ensures that the holomorphic function
\be
h(\bar z)=\frac{\tilde\omega(\bar z)e^{-\frac{\pi}{2}z\bar z}}{\omega(\bar z)e^{-\frac{\pi}{2}z\bar z}}\,,
\label{eq:A4}
\ee
has no poles as all zeroes of $\omega(\bar z)$ give rise to at most single poles, at least compensated by a zero of the function $\tilde\omega$. The extra exponential is added
to \eq{eq:A4} in order
to make the denominator finite and periodic in the whole plane. Moreover, one can easily check that
this additional exponential
makes the numerator double periodic, with the periods twice those of the denominator. Therefore, the function $\vert h(\bar z)\vert$ has at least the same periodicity as $\tilde\omega(\bar z)$.

We have now a holomorphic, periodic and finite function $h(\bar z)$, which must therefore be a constant.
The normalization parameter $N$ in Eq.~\eq{w2} is then determined from the condition $h(\bar z) = 1$ at any point $\bar z$.
For example, at the saddle point  $\bar z_h=-\frac{1+\tau}{2}$ one finds:
	 	\be
	 	\omega(\bar z_h)e^{-\frac{\pi}{2}\vert z_h\vert^2}
{\biggl|}_{\bar z_h=-\frac{1+\tau}{2}}
=-e^{\frac{i\pi}{4}}\frac{\eta^2(\frac{1+\tau}{2})}{\eta(1+\tau)},
	 	\ee
where $\eta$ is the Dedekind eta function.
Then the prefactor $N$ can be evaluated numerically:
	 	\be
	 	N=3.1825\ldots
	 	\ee

Thus, we conclude that $\tilde\omega(\bar z)\equiv \omega(\bar z)$, and
\beqn
		\theta_1(\bar z, e^{i\pi\tau}) {\biggl|}_{\tau=e^{i\frac{\pi}{3}}} & = & N e^{\frac{i\pi}{8}}\sum_{m,n} (\bar z-\bar X_{m,n})
		\label{eq:theta1:equality}\\
		& & \times e^{-\frac{\pi}{2}\vert z- X_{m,n}\vert^2+\frac{\pi}{2}\left(\bar z X_{m,n}-z\bar X_{m,n}\right)}, \nonumber
\eeqn
is an alternative convenient expansion of the $\theta_1$ function which enter the solution in Eq.~(\ref{zetaeq}).
The validity of Eq.~\eq{eq:theta1:equality} can also be checked numerically.


\begin{thebibliography}{99}
\bibitem{ref:footnote1}
Another solution, $\omega(\bar z)=\theta_3\left(-i \nu\bar z,q\right)\exp\{- \pi \bar z^2/2 \}$, is essentially identical to the solution (\ref{def:omega}) modulo a translation and rotation in the transverse plane.
\end{thebibliography}
\end{document}